\documentclass{article}

\usepackage[utf8]{inputenc}
\usepackage{amsmath, amssymb}
\usepackage{amsthm}
\newtheorem{proposition}{Proposition}
\usepackage{rotating}
\usepackage{graphicx}
\everymath{\displaystyle}

\usepackage{natbib}

\title{The order book as a queueing system: average depth and influence of the size of limit orders}
\author{Ioane Muni Toke
\\ ERIM, University of New Caledonia
\\ BP R4 98851 Noumea CEDEX, New Caledonia
\\ \tt{ioane.muni-toke@univ-nc.nc}
\\ \vspace{6pt}
\\ Applied Maths Laboratory, Chair of Quantitative Finance
\\ Ecole Centrale Paris 
\\ Grande Voie des Vignes, 92290 Châtenay-Malabry, France
\\ \tt{ioane.muni-toke@ecp.fr}
}
\date{}

\begin{document}
\maketitle
\begin{abstract}
In this paper, we study the analytical properties of a one-side order book model in which the flows of limit and market orders are Poisson processes and the distribution of lifetimes of cancelled orders is exponential.
Although simplistic, the model provides an analytical tractability that should not be overlooked.
Using basic results for birth-and-death processes, we build an analytical formula for the shape (depth) of a continuous order book model which is both founded by market mechanisms and very close to empirically tested formulas. We relate this shape to the probability of execution of a limit order, highlighting a law of conservation of the flows of orders in an order book.
We then extend our model by allowing random sizes of limit orders, hereby allowing to study the relationship between the size of the incoming limit orders and the shape of the order book. Our theoretical model shows that, for a given total volume of incoming limit orders, the less limit orders are submitted (i.e. the larger the average size of these limit orders), the deeper is the order book around the spread. This theoretical relationship is finally empirically tested on several stocks traded on the Paris stock exchange.
\end{abstract}

\section{Introduction}

The limit order book is central to modern electronic financial markets. For any given stock or financial product, this structure centralizes at any time the bid and offer of all traders.
Orders submitted by traders to a given market place may be of very different types, depending on the trading rules of the market place, the security traded, etc. However, they can generally be categorized in three main types:
\begin{itemize}
	\item limit orders: buy orders submitted at a price lower than the current ask price and sell orders submitted at a price higher than the current bid price;
	\item market orders: buy and sell orders to be filled instantly, whatever the price;
	\item cancel orders: cancellation of a previously submitted limit order now standing in the order book, i.e. not yet executed.
\end{itemize}
The combination of the flows of orders of these three types leads to the dynamics of the order book. The characteristics of the order book -- among which the shape and the resilience -- roughly define what is commonly understood in market microstructure literature as the liquidity \citep[see e.g.][]{Foucault2005}.
We call "shape" of the order book the function which for any price gives the number of shares standing in the order book at that price. We call "cumulative shape" up to price $p$ the total quantity offered in the order book between the best limit and price $p$. These quantities measure what is sometimes known as the depth of the order book. In a probabilistic model of the order book in which the random variable describing the (cumulative) shape admits a stationary distribution, its expectation with respect to this stationary distribution will be simply called the average (cumulative) shape.

Understanding the average shape of an order book and its link to the order flows is not straightforward.
\cite{Biais1995} is a pioneer empirical study of the 40 largest capitalizations of the Paris Bourse (using 1991 data). It is observed that, at least for the first limits, the shape of the order book is increasing away from the spread.
A decade later, and still on the Paris Bourse, \cite{Bouchaud2002} and \cite{Potters2003} propose improved results. It is observed (using 2001 data) that "the average order book has a maximum away from the current bid/ask, and a tail reflecting the statistics of the incoming orders" (hump-shaped limit order book). A power law distribution is suggested for modelling the decrease of the flow of limit orders away from the spread.
Furthermore, \cite{Bouchaud2002} provides an analytical approximation to fit the average shape of the order book, that depends on the incoming flow of limit orders.
\cite{Gu2008} confirms the hump-shaped form of the order book by an empirical study of liquid Chinese stocks (using 2003 data), but finds an exponential decrease of the average shape as the price increases. \cite{Russell2010} also observes the hump-shaped for the US Google stock (using 2005 data).

More recently, research about optimal execution in an order book and order book simulations has risen interest in understanding the shape of orders books. 
Optimal execution problems deals with the shape of the order book, since a proxy for the price impact of a trade is the inverse function of the cumulative shape of the order book \citep[see e.g.][]{Bouchaud2009}. Therefore, these problems need an order book shape as input: block-shape has been used in toy models, general shapes have recently been considered \citep{Obizhaeva2012,Predoiu2011,Alfonsi2010}. 
As for order book simulations, once the flows of orders have been defined, checking the simulated shape is important in assessing that the model is realistic. A log-normal shape of the order book has been suggested \citep{Preis2006}.

Among recent applied probability works answering to this new need for order book modelling, let us mention the Markovian model of \cite{Yudovina2012} or the model based on measure-valued processes in \cite{Lakner2013}. Note however that these works do not include any cancellation mechanism, limiting their immediate practical applicability.

An order book is primarily a complex queueing system. Recent studies \citep{Cont2010,Cont2011} build on this observations and use classical tools from queueing theory such as the computation of Laplace-Stieljes transforms of distributions of interest. \cite{Cont2010} observes that "the average profile of the order book displays a hump [\ldots] that does not result from any fine-tuning of model parameters or additional ingredients such as correlation between order flow and past price moves", but no explicit link between the average shape and the order flows is made. Our paper hopefully closes this gap.
We start with the simplest queueing system as an order book: we consider a one-side limit order book model in which limit orders and market orders occur according to Poisson processes. It is also assumed that all limit orders are submitted with an exponentially distributed lifetime, i.e. they are cancelled if they have not been executed some exponential time after their submission. This starting model is discrete in the sense that the price is an integer, expressed in number of ticks.

In section \ref{sec:OBQueue}, this model is described and we show recalling simple results from queueing theory that this system admits an analytically computable expected size associated to its stationary distribution. We then develop this into a continuous model by replacing the finite set of Poisson processes describing the arrival of limit orders with a spatial Poisson process. The main interest of the continuous version of the model is to allow easier analytical manipulation and to describe the frequent case in which the tick size is small compared to standard variations of prices.
We obtain an analytical formula for the average shape of the continuous order book, that depends explicitly on the arrival rates of limit, market and cancellation orders. We interpret this shape in terms of the  probability that a submitted limit order is cancelled (as opposed to executed), introducing a law of conservation of the flows of orders in a Markovian order book.

In section \ref{sec:Comparison}, we compare our analytical formula for the shape of the order book to the numerical investigations of \cite{Smith2003}, and then to the only (to our knowledge) existing analytical formula for the shape of an order book available in the literature, proposed by \cite{Bouchaud2002}.

Section \ref{sec:ModelVolumes} extends the previous model by allowing for random sizes of limit orders. We update our analytical formulas in the special case in which the sizes of limit orders are assumed to be geometrically distributed.
Thanks to this model, we investigate in section \ref{sec:InfluenceOrderSize} the role of the size of limit orders in an order book, providing insights on its influence on the order book shape. 
Few empirical literature provides insights on the factors that make the average shape of order books vary.
\cite{Beltran2009} analyses the variations of order books on the German Stock Exchange (2004 data) through a principal component analysis and identifies two main factors that respectively shift and rotate the shape of the order book.
\cite{Naes2006} investigates the slope of the order book around the spread as a function of the traded volume on 108 Norwegian stocks (2001 data) and, among other results, exhibit a positive relation between the order book depth around the spread and the trading activity: high trading activity tends to deepen the first limits of an order book.
Our model suggests an influence of the relative size of limit orders, that we empirically test using 2010 trading data on the Paris Stock Exchange. Section \ref{sec:Conclusion} concludes this work.

\section{A link between the flows of order and the shape of an order book}
\label{sec:OBQueue}

\subsection{The basic queueing system}
\label{subsec:DiscreteModel}

The aim of this section is to present the basic one-side order book model, discuss the relevance of its assumptions and recall some results from queueing theory in the context of this order book model.
Let us consider a one-side order book model, i.e. a model in which all limit orders are ask orders, and all market orders are buy orders.
Bid price is assumed to be constantly equal to zero, and consequently spread and ask price are identically equal. From now on, this quantity will be simply refered to as the price.
Let $p(t)$ denote the price at time $t$. $\{p(t), t\in[0\in\infty)\}$ is a continuous-time stochastic process with value in the discrete set $\{1,\ldots,K\}$. In other words, the price is given in number of ticks. 
Let $\Delta$ be the tick size, such that the price range of the model in currency is actually $\{\Delta,\ldots,K\Delta\}$.
For realistic modelling and empirical fitting performance, one may assume that the maximum price $K$ is chosen very large, but in fact it will soon be obvious that this upper bound does not affect in any way the order book for lower prices. 
For all $i\in\{1,\ldots,K\}$, (ask) limit orders at price $i$ are submitted according to a Poisson process with parameter $\lambda_i$. These processes as assumed to be mutually independent, so that the number of orders submitted at prices $q\ldots,r$ is a Poisson process with parameter $\lambda_{q\rightarrow r}$ defined as $\lambda_{q\rightarrow r}= \sum_{i=q}^r\lambda_i$.
All limit orders standing in the book may be cancelled. It is assumed that the time intervals between submission and cancellation form a set of mutually independent random variables identically distributed according to an exponential distribution with parameter $\theta>0$.
Finally, (buy) market orders are submitted at random times according to a Poisson process with parameter $\mu$. Note that all orders are assumed to be of unit size. This restriction will be dropped in sections \ref{sec:ModelVolumes} et sq.

Let us comment here on the model specification.
Firstly, paremeter $\theta$ is difficult to grasp empirically, and the use, for our analytical purposes, of the unconditional exponential distribution is rough. \cite{Challet2001} finds a power-law distribution for this quantity, confirmed in \cite{Chakraborti2011}. It seems plausible that the lifetime of limit orders that are cancelled in an order book varies with the distance between the order price and the current ask price, among many other parameters. \cite{Mike2008} outperforms the basic exponential model with an empirically-defined three-parameter cancellation model, too complex to be analytically used here.
Secondly, as for the limit and market orders, several empirical studies show that the Poisson processes do not reflect the complexity of the order flows. Complex point processes have been suggested to model the interactions between the order flows, and the consequent resilience of the order book \citep[e.g.][]{Large2007,Toke2011}.
We however focus on a order flows as Poisson processes in order to keep as much analytical tractability as possible.
Thirdly, focusing on a one-side limit order book answers to the same need for simplicity. Such models are often considered in optimal execution papers \citep[e.g.][]{Obizhaeva2012,Predoiu2011,Alfonsi2010} or in empirical papers \citep[e.g.][]{Bouchaud2002} where a bid/ask symmetry is assumed.
Our specification is equivalent to a two-side model with an infinite-volume limit bid order standing in the book at price $0$. This situation could be found on a market where several sellers would make continuous-time offers that may be cancelled to a single-buyer that alone decides when to buy.
Note also that our toy model is a half-book of the double-side book studied by \cite{Cont2010}.
Finally, despite these very simple specifications, we will obtain general flexible order shapes that are both empirically valid and founded by market mechanisms.

Let $\{L^{1\to k}(t),t\in[0,\infty)\}$ be the stochastic process representing the number of limit orders at prices $1,\ldots,k$ standing in the order book at time $t$.
$L^{1\to k}$ is thus the cumulative shape of the order book in our model.
$L^{1\to k}$ can be viewed as a birth-death process with birth rate $\lambda_{1\to k}$ et death rate $\mu+n\theta$ in state $n$ ; it may equivalently be viewed as the size of a $M/M/1+M$ queueing system with arrival rate $\lambda_{1\to k}$, service rate $\mu$ and reneging rate $\theta$ (see e.g. \cite[Chapter XVII]{Feller1968} or \cite[Chapter 8]{Bremaud1999} among many textbook references). This queueing system will now be refered to as the $1\to k$ queueing system.
$L^{1\to k}$ admits a stationary distribution $\pi_{1\to k}(\cdot)$ as soon as $\theta>0$.
Using the formalism of the infinitesimal generator:
\begin{equation}
	A=\left(\begin{array}{cccccc}
	-\lambda_{1\to k} & \lambda_{1\to k} & 0 & 0 & 0 & \ldots
	\\
	\mu+\theta & -(\lambda_{1\to k}+\mu+\theta) & \lambda_{1\to k} & 0 & 0 & \ldots
	\\
	0 & \mu+2\theta & -(\lambda_{1\to k}+\mu+2\theta) & \lambda_{1\to k} & 0 & \ldots
	\\
	\vdots & \vdots & \ddots & \ddots & \ddots & \ddots
	\end{array}
	\right),
\end{equation}
this stationary probability $\pi_{1\to k}$ is classically obtained and written for all $n\in\mathbb N^*$:
\begin{equation}
	\pi_{1\to k}(n) = \pi_{1\to k}(0) \prod_{i=1}^n \frac{\lambda_{1\to k}}{\mu+i\theta},
\end{equation}
and setting $\sum_{n=0}^\infty \pi(n) = 1$ gives:
\begin{equation}
	\pi_{1\to k}(0) = \left(\sum_{n=1}^\infty\prod_{i=1}^n \frac{\lambda_{1\to k}}{\mu+i\theta}\right)^{-1}.
\end{equation}
Introducing the normalized parameters $\nu_{1\to k} = \frac{\lambda_{1\to k}}{\theta}$ and $\delta = \frac{\mu}{\theta}$,
and after some simplifications, we write for all $n\in\mathbb N$:
\begin{equation}
	\pi_{1\to k}(n) = \frac{e^{-\nu_{1\to k}} \nu_{1\to k}^{\delta}}{\delta\Gamma_{\nu_{1\to k}}(\delta)} \prod_{i=1}^n \frac{\nu_{1\to k}}{i+\delta},
	\label{eq:Pi1Tokn}
\end{equation}
where $\Gamma_y$ is the lower incomplete version of the Euler gamma function:
\begin{equation}
	\Gamma_y:\mathbb R_+\to\mathbb R, x\mapsto\int_0^y t^{x-1}e^{-t}dt.
\end{equation}
Now, since the price is equal to $k$ if and only if the "$1\to k-1$" queueing system is empty and the "$1\to k$" system is not, we may state that in the one-side order book model, the stationary distribution $\pi_p$ of the price $p$ is written:
\begin{equation}
	\label{eq:DiscretePrice1}
	\pi_p(1) = 1-\frac{e^{-\nu_{1\to k}} \nu_{1\to k}^{\delta}}{\delta\Gamma_{\nu_{1\to k}}(\delta)},
\end{equation}
and for all $k\in\{2,\ldots,K\}$,
\begin{equation}
	\label{eq:DiscretePrice2+}
	\pi_p(k) = \frac{e^{-\nu_{1\to k-1}} \nu_{1\to k-1}^{\delta}}{\delta\Gamma_{\nu_{1\to k-1}}(\delta)} - \frac{e^{-\nu_{1\to k}} \nu_{1\to k}^{\delta}}{\delta\Gamma_{\nu_{1\to k}}(\delta)}.
\end{equation}
Using previous results, the average size $\mathbf E[L^{1\to k}]$ of the "$1\to k$" queueing system is easily computed. From equation \eqref{eq:Pi1Tokn}, we can write after some simplifications:
\begin{equation}
	\label{eq:DiscreteCumulativeShape}
	\mathbf E[L^{1\to k}] = \nu_{1\to k}-\frac{\Gamma_{\nu_{1\to k}}(1+\delta)}
	{\Gamma_{\nu_{1\to k}}(\delta)}.
\end{equation}
Let $L^k=L^{1\to k}- L^{1\to k-1}$ be the number of orders in the book at price $k\in\{1,\ldots,K\}$. Then the average shape of the order book at price $k$ is obviously :
\begin{equation}
	\label{eq:DiscreteShape}
	\mathbf E[L^k] = \nu_k - \left(\frac{\Gamma_{\nu_{1\to k}}(1+\delta)}{\Gamma_{\nu_{1\to k}}(\delta)} - \frac{\Gamma_{\nu_{1\to k-1}}(1+\delta)}{\Gamma_{\nu_{1\to k-1}}(\delta)}\right).
\end{equation}

\subsection{A continuous extension of the basic model}

In order to facilitate the comparison with existing results, we derive a continuous version of the previous toy model.
Price is now assumed to be a positive real number. Mechanisms for market orders and cancellations are identical: unit-size market orders are submitted according to a Poisson process with rate $\mu$, and standing limit orders are cancelled after some exponential random time with paramter $\theta$.
As for the submission of limit orders, the mechanism is now slightly modified: since the price is continuous, instead of a finite set of homogeneous Poisson processes indexed by the number of ticks $k\in\{1,\ldots,K\}$, we now consider a spatial Poisson process on the positive quadrant $\mathbb R_+^2$.
Let $\lambda(p,t)$ be a non-negative function denoting the intensity of the spatial Poisson process modelling the arrival of limit orders, the first coordinate representing the price, the second one the time \citep[see e.g.][Chapter 12 for a textbook introduction on the construction of spatial Poisson processes]{Privault2013}. 
As in the discrete case, this process is assumed to be time-homogeneous, and it is hence assumed that price and time are separable. Let $h_\lambda:\mathbb R_+\to\mathbb R_+$ denote the spatial intensity function of the random events, i.e. limit orders. Then, $\lambda(p,t)= \alpha h_\lambda(p)$ is the intensity of the spatial Poisson process representing the arrival of limit orders.

We recall that in this framework, for any $p_1<p_2\in[0,\infty)$, the number of limit orders submitted at a price $p\in[p_1,p_2]$ is a homogeneous Poisson process with intensity $\int_{p_1}^{p_2} \lambda(p,t)\,dp = \alpha \int_{p_1}^{p_2} h_\lambda(p)\,dp$. Furthermore, if $p_1<p_2<p_3<p_4$ on the real positive half-line, then the number of limit orders submitted in $[p_1,p_2]$ and $[p_3,p_4]$ form two independent Poisson processes.

Now, let $L([0,p])$ be the random variable describing the cumulative size of our new order book up to price $p\in\mathbb R_+$. Given the preceding remarks, $L([0,p])$ is, as in the previous section, the size of a $M/M/1+M$ queueing system with arrival rate $\alpha \int_0^p h_\lambda(u)\,du$, service rate $\mu$ and reneging rate $\theta$. Using the results of section \ref{subsec:DiscreteModel}, we obtain from equation \eqref{eq:DiscreteCumulativeShape}:
\begin{equation}
	\label{eq:ContCumShape}
	\mathbf E[L([0,p])] = \int_0^p h(u) \,du - f\left(\delta, \int_0^p h(u) \,du \right),
\end{equation}
where we have defined $h(u)=\frac{\alpha h_\lambda(u)}{\theta}$ and : 
\begin{equation}
	\label{eq:functionf}
	f(x,y)=\frac{\Gamma_{y}(1+x)}{\Gamma_{y}(x)}.
\end{equation}
From now on, $H(p) = \int_0^p h(u) \, du$ will be the (normalized) arrival rate of limit orders up to price $p$, and $B(p)=\mathbf E[L([0,p])]$ will be the average cumulative shape of the order book up to price $p$. Then $b(p)=\frac{dB(p)}{dp}$ will be the average shape of the order book (per price unit, not cumulative). Straightforward differentiation of equation \eqref{eq:ContCumShape} and some terms rearrangements lead to the following proposition.
\begin{proposition}
\label{propo:ContShape}
In a continuous order book with homogeneous Poisson arrival of market orders with intensity $\mu$, spatial Poisson arrival of limit orders with intensity $\alpha h_\lambda(p)$, and exponentially distributed lifetimes of non-executed limit orders with parameter $\theta$, the average shape of the order book $b$ is computed for all $p\in[0,\infty)$ by:
\begin{equation}
	\label{eq:ContShape}
	b(p) = h(p) \left[ 1 - \delta \left(g_\delta \circ H\right)(p) \left[ 1 - 
	\delta H^{-1}(p) \left[ 1 - \left(g_\delta \circ H\right)(p) \right]
   \right]\right],
\end{equation}
where
\begin{equation}
	\label{eq:defigdelta}
	g_\delta (y) = \frac{e^{-y} y^{\delta }}{\delta\Gamma_y(\delta)}.
\end{equation}
\end{proposition}

Let us give a few comments on the average shape we obtain. Firstly, note that by identification to equation \eqref{eq:Pi1Tokn}, observing that $\pi_{1\to k}(0) = g_\delta(\nu_{1\to k})$ in the discrete model, $g_\delta(H(p))$ is to be interpreted as the probability that the order book is empty up to a price $p$. Secondly, note that letting $\delta\to 0$ in equation \eqref{eq:ContShape} gives $b(p)\to h(p)$ (cf. $\lim_{\delta\to 0} g_\delta(y) = e^{-y}$). Indeed, if there were no market orders, then the average shape of the order book would be equal to the normalized arrival rates.
Thirdly, as $p\to\infty$, we have $b(p)\sim k\,h(p)$ for some constant $k$.
This leads to our main comment, which we state as the following proposition.
\begin{proposition}
The shape of the order book $b(p)$ can be written as:
\begin{equation}
	b(p) = h(p)C(p),
\end{equation}
where $C(p)$ is the probability that a limit order submitted at price $p$ will be cancelled before being executed.
\end{proposition}
This proposition translates a \emph{law of conservation of the flows of orders}: the shape of the order book is exactly the fraction of arriving limit orders that will be cancelled. The difference between the flows of arriving limit orders and the order book is exactly the fraction of arriving limit orders that will be executed.

The proof is straightforward. Indeed, in the $1\to k$ queueing system, the average number of limit orders at price $k$ that are cancelled per unit time is $\theta\mathbf E[L^k]$ ($\theta \mathbf E[L^{1\to k}]$ is the reneging rate of $1\to k$ queue using queueing system vocabulary). Therefore, the fraction of cancelled orders at price $k$ over arriving limit orders at price $k$, per unit time, is
$C_k=\frac{\theta\mathbf E[L^k]}{\lambda_k}$.
Using equation \eqref{eq:DiscreteShape} and some straightforward computations, the fraction $C_k$ of limit orders submitted at price $k$ which are cancelled is:
\begin{equation}
	C_k = 1 - \frac{\delta}{\nu_k} \left( g_\delta(\nu_{1\to k-1}) - g_\delta(\nu_{1\to k}) \right).
\end{equation}
Therefore, in the continuous model up to price $p\in\mathbb R_+$ with average cumulative shape $B(p)$, the fraction of limit orders submitted at a price in $[p,p+\epsilon]$ which are cancelled is written:
\begin{equation}
	1-\frac{\delta}{H(p+\epsilon)-H(p)} \left( g_\delta(H(p)) - g_\delta(H(p+\epsilon)) \right).
\end{equation}
By letting, $\epsilon\to 0$, we obtain that the fraction $C(p)$ of limit orders submitted at price $p\in\mathbb R_+$ which are cancelled is :
\begin{equation}
	C(p) = 1-\delta g'_\delta (H(p)) = 1-\delta  (g_\delta\circ H)(p) 
	\left( 1- \frac{\delta}{H(p)} (1 - (g_\delta\circ H)(p) )
	\right),
\end{equation}
which gives the final result.

This law of conservations of the flows of orders explains the relationship between the shape of the order book and the flows of arrival of limit orders. For high prices, two cases are to be distinguished. On the one hand, if the total arrival rate of limit orders is a finite positive constant $\alpha$ (for example when $h_\lambda$ is a probability density function on $[0,+\infty)$, in which case $\lim_{p\to+\infty} H(p)=\int_0^\infty \lambda(u,t)\,du=\alpha\in\mathbb R_+^*$), then, the proportionality constant between the shape of the order book $b(p)$ and the normalized limit order flow $h(p)$ is, as $p\to+\infty$, $C_\infty$ defined as:
\begin{equation}
	\label{eq:LimitInfinityProbaCancellation}
	C_\infty = \lim_{p\to\infty} C(p) = 1-\delta g_\delta(\alpha)\left(1-\frac{\delta}{\alpha}\left(1-g_\delta(\alpha)\right)\right)<1.
\end{equation}
In such as case, the shape of the order book as $p\to+\infty$ is proportional to the normalized rate of arrival of limit orders $h(p)$, but not equivalent. The fraction of cancelled orders does not tend to $1$ as $p\to+\infty$, i.e. market orders play a role even at high prices.
On the other hand, in the case where $\lim_{p\to+\infty} H(p)=\infty$, then very high prices are not reached by market orders, and the tail of the order book behaves exactly as if there were no market orders: $b(p)\sim h(p)$ as $p\to+\infty$.
We may remark here that there exists an empirical model for the probability of execution $F(p)=1-C(p)$ in \cite{Mike2008}. In this model, it is assumed to be the complementary cumulative distribution function of a Student distribution with parameter $s=1.3$. As such, it is decreasing towards $0$ as $p^{-s}$. In our model however, it is exponentially decreasing, and, in view of the previous discussion, does not necessarily tends towards $0$.

\section{Comparison to existing results on the shape of the order book}
\label{sec:Comparison}

The model presented in section \ref{sec:OBQueue} belongs to the class of "zero-intelligence" Markovian order book models: all order flows are independent Poisson processes. Although very simple, it turns out it replicates the shapes of the order book usually obtained in previous empirical and numerical studies, as we will now see.

\subsection{Numerically simulated shape in \cite{Smith2003}}
\label{subsec:CompSmith}

A first result on the shape of the order book is provided in \cite{Smith2003}, on figures 3a) and 3b). These figures are obtained by numerical simulation of an order book model very similar to the one presented in section \ref{sec:OBQueue}, where all order flows are Poisson processes: market orders are submitted are rate $\mu_S$ with size $\sigma_S$, limit orders are submitted with the same size at rate $\alpha_S$ per unit price on a grid with tick size $dp_S$, and all orders are removed randomly with constant probability $\delta_S$ per unit time. (We have indexed all variables with an $S$ to differentiate them from our own notations).
Figures 3a) and 3b) in \cite{Smith2003} are obtained for different values of a "granularity" parameter $\epsilon_S\propto\frac{\delta_S\sigma_S}{\mu_S}$. It is observed that, when $\epsilon_S$ gets larger, the average book becomes deeper close to the spread, and thinner for higher prices.

\begin{figure}
\begin{center}
\begin{tabular}{c}
\includegraphics[scale=1]{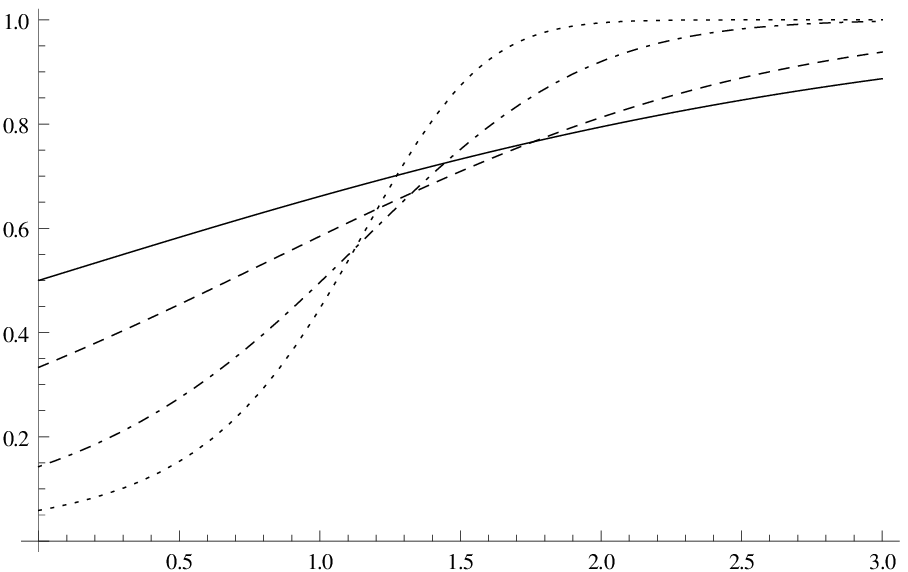}
\\
\includegraphics[scale=1]{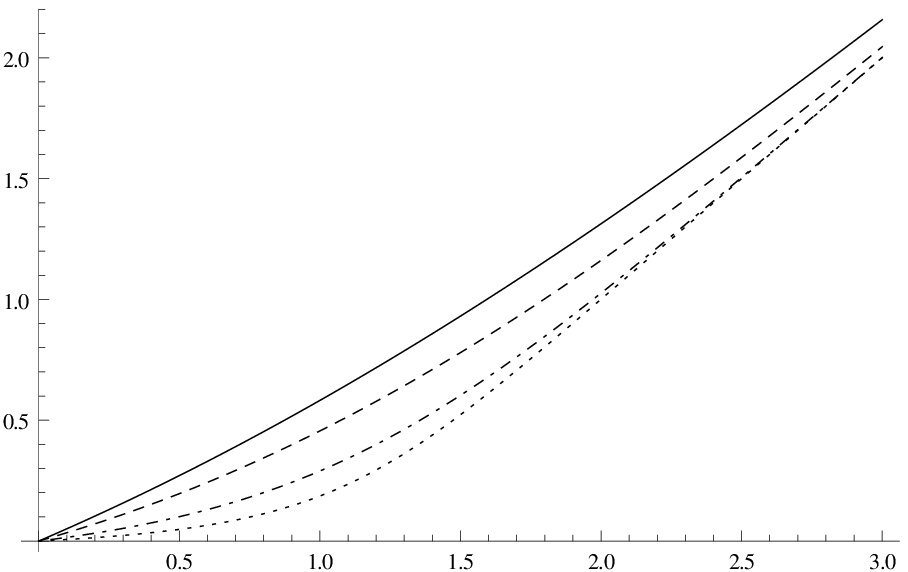}
\end{tabular}
\end{center}
\caption{Shape (top panel) and cumulative shape (bottom panel) of the order book computed using equations \eqref{eq:ContShape} and \eqref{eq:ContCumShape} with $h(p)=\alpha$, $\alpha=8$ and $\delta=1$ (full line), $\delta=2$ (dashed), $\delta=6$ (dotdashed), $\delta=16$ (dotted). Note that results are scaled on the same dimensionless axes used in \cite{Smith2003}.}
\label{fig:OBShapeCumShapeUnitVolumeVaryingDelta}
\end{figure}

Using our own notations, $\epsilon_S$ actually reduces to $\frac{1}{\delta}$, i.e. the inverse of the normalized rate of arrival of limit orders.
Using \cite{Smith2003}'s assumption that limit orders arrive at constant rate $\alpha_S$ per unit price and unit time, we obtain in our model $\lambda(p,t)=\alpha_S$, i.e. $H(p)=\alpha_S p$.
On figure \ref{fig:OBShapeCumShapeUnitVolumeVaryingDelta}, we plot the average shapes and cumulative shapes of the order book given at equations \eqref{eq:ContShape} and \eqref{eq:ContCumShape} with this $H$.
It turns out that when $\delta$ varies, our basic model indeed reproduces precisely figure 3a) and 3b) of \cite{Smith2003}.
Therefore we are able to analytically describe the shapes that were only numerically obtained. These shapes can be straightforwardly obtained with different regimes of market orders in our basic model: when the arrival rates of market orders increases (i.e. when $\epsilon_S$ increases), all other things being equal, the average shape of the order book is thinner for lower prices.

\subsection{Empirical and analytical shape in \cite{Bouchaud2002}}
\label{subsec:CompBouchaud}

We now give two more examples of order book shapes obtained with equation \eqref{eq:ContShape} of proposition \ref{propo:ContShape}. 
We successively consider two types of normalized intensities of arrival rates of limit orders: 
\begin{itemize}
	\item exponentially decreasing with the price: $h(u)=\alpha\beta e^{-\beta u}$ ;
	\item power-law decreasing with the price: $h(u)=\alpha(\gamma-1)(1+u)^{-\gamma}, \gamma>1$.
\end{itemize}
The first case is the one observed on Chinese stocks by \cite{Gu2008}.
The second case is the one suggested in an empirical study by \cite{Bouchaud2002}, in which $\gamma\approx1.5-1.7$.
Moreover, the latter paper provides the only analytical formula previoulsy proposed (to our knowledge) linking the order flows and the average shape of a limit order book: \cite{Bouchaud2002} derives an analytical formula from a zero-intelligence model by assuming that the price process is diffusive with diffusion constant $D$.
Using our notations, their average order book, denoted here $b_{BP}$, is for any $p\in(0,\infty)$:
\begin{equation}
	\label{eq:BPShape}
	b_{BP}(p) \propto e^{-\sigma p} \int_0^p h_\lambda(u)\sinh(\sigma u)\,du
	+ \sinh(\sigma p)\int_p^{\infty}h_\lambda(u)e^{-\sigma u}\,du,
\end{equation}
with $\sigma = \sqrt{\frac{D}{2\theta}}$.

\begin{figure}
\begin{center}
\includegraphics[scale=1]{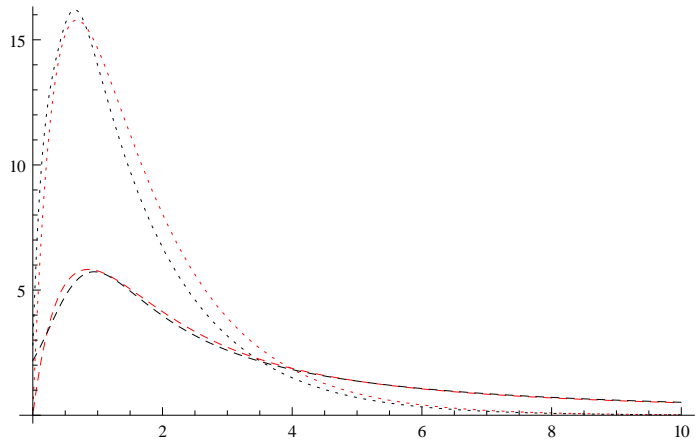}
\caption{Comparison of the shapes of the order book in our model (black curves) and using the formula proposed by \cite{Bouchaud2002} (red curves), for three cases of limit order arrival rates, exponential (dotted), power-law (dashed). For the red curves, coefficients $D$ is respectively $0.30$ and $0.40$.}
\label{fig:OBShapesCompToBouchaud}
\end{center}
\end{figure}

We plot the shape of \cite{Bouchaud2002} for the two types of normalized arrival rates of limit orders previously mentionned. $D$ is the only variable that is not directly available in our model. In all cases, we use as a proxy the standard deviation of the price in our model, obtained by numerical simulation\footnote{We use the numerical simulations for the ease of use, but one may also analytically compute the standard deviation of the price in the model using equations \eqref{eq:DiscretePrice1} and \eqref{eq:DiscretePrice2+}.}. Note also that the formula \eqref{eq:BPShape} is defined up to a multiplicative constant that we arbitrarily (and roughly) fix such that the maximum offered with respect to the price in our model approximately corresponds to the maximum of equation \eqref{eq:BPShape}.
Results are plotted in figure \ref{fig:OBShapesCompToBouchaud}, and numerical values given in caption.
It turns out that our model \eqref{eq:ContShape} and equation \eqref{eq:BPShape} provide remarkably similar order book shapes.
Since equation \eqref{eq:BPShape} has been successfully tested with empirical data in \cite{Bouchaud2002}, figure \ref{fig:OBShapesCompToBouchaud} provides a good hint that the shape \eqref{eq:ContShape} should provide very good empirical fittings as well.
As $p\to\infty$, both formulas lead to a shape $b(p)$ decreasing as the arrival rate of limit orders $h(p)$, which was already observed in \cite{Bouchaud2002}, and discussed here in section \ref{sec:OBQueue}. The main difference between the shapes occurs as $p\to 0$.
Equation \eqref{eq:BPShape} imposes that $b_{BP}=0$, whereas the result \eqref{eq:ContShape} allows a more flexible behaviour with $b(0)=\frac{h(0)}{1+\delta}$, a quantity that depends on the three types of order flows. The fact that $b_{BP}$ does not depend directly on the flows of market orders, but only through the constant $D$ is probably the source the different behaviours in that case.

\section{A model with varying sizes of limit orders}
\label{sec:ModelVolumes}

We now allow for random sizes of limit orders in our model. As in section \ref{sec:OBQueue}, we start by describing the basic model as a queueing system, and then extend it to the case of a continuous price.

Let us recall that we deal with a one-side order book model, i.e. a model in which all limit orders are ask orders, and all market orders are bid orders.
Let $p(t)$ denote the price at time $t$. $\{p(t), t\in[0\in\infty)\}$ is a continuous-time stochastic process with value in the discrete set $\{1,\ldots,K\}$, i.e. the price is given in number of ticks. 
For all $i\in\{1,\ldots,K\}$, (ask) limit orders at price $i$ are submitted according to a Poisson process with parameter $\lambda_i$. These processes as assumed to be mutually independent, so that the number of orders submitted at prices between $q$ and $r$ (included) is a Poisson process with parameter $\lambda_{q\rightarrow r}$ defined as $\lambda_{q\rightarrow r}= \sum_{i=q}^r\lambda_i$.

The contribution of this section is to allow for random sizes of limit orders, instead of having unit-size limit orders as in the basic model of section \ref{sec:OBQueue}. We assume that all the sizes of limit orders are independent random variables. We also assume that the sizes of limit orders submitted at a given price are identically distributed, but we allow this distribution to vary depending on the price. For a given price $k\in\mathbb N^*$, let $g^k_n$, $n\in\mathbb N^*$, denote the probability that a limit order at price $k$ is of size $n$. Let $\overline{g^k}$ denote the mean size of a limit order at price $k$, which is assumed to be finite.
It is a well known property of the Poisson process to state that the rate of arrival of limit orders of size $n$ at price $i$ is $\lambda_i g^i_n$, hence the rate of arrival of limit orders of size $n$ with a price lower or equal to $k$ is $\sum_{i=1}^k \lambda_i g^i_n$.
Similarly, the probability that a limit order with a price lower or equal to $k$ is of size $n$ is $g^{1\to k}_n = \sum_{i=1}^k \frac{\lambda_i}{\lambda_{1\to k}} g^i_n$. Let $\overline{g^{1\to k}} =  \sum_{i=1}^k \frac{\lambda_i}{\lambda_{1\to k}} \overline{g^i}$ denote the mean size of a limit order with price up to $k$.

Mechanism for cancellation is unchanged: all limit orders standing in the book may be cancelled. Note however that a limit order is not cancelled all at once, but unit by unit (i.e. share by share). It is assumed that the time intervals between the submission of a limit order and the cancellation of one share of this order form a set of mutually independent random variables identically distributed according to an exponential distribution with parameter $\theta>0$.
Finally, (buy) market orders are submitted at random times according to a Poisson process with parameter $\mu$. All market orders are assumed to be of unit size.

As in section \ref{sec:OBQueue}, let $\{L^{1\to k}(t),t\in[0,\infty)\}$ be the stochastic process representing the number of limit orders at prices $1,\ldots,k$ standing in the order book at time $t$.$L^{1\to k}$ is thus the cumulative shape of the order book in our model.
It can be viewed as the size of a $M^X/M/1+M$ queueing system with bulk arrival rate $\lambda_{1\to k}$, bulk volume distribution $(g^{1\to k}_n)_{n\in\mathbb N^*}$, service rate $\mu$ and reneging rate $\theta$ (see e.g. \cite{Chaudhry1983} for queueing systems with bulk arrivals).
The infinitisemal generator of the process $L^{1\to k}$ is thus written:
\begin{equation}
	\left(\begin{array}{cccccc}
	-\lambda_{1\to k} & \lambda_{1\to k}g^{1\to k}_1 & \lambda_{1\to k}g^{1\to k}_2 & \lambda_{1\to k}g^{1\to k}_3 & \lambda_{1\to k}g^{1\to k}_4 & \ldots
	\\
	\mu+\theta & -(\lambda_{1\to k}+\mu+\theta) & \lambda_{1\to k}g^{1\to k}_1 & \lambda_{1\to k}g^{1\to k}_2 & \lambda_{1\to k}g^{1\to k}_3 & \ldots
	\\
	0 & \mu+2\theta & -(\lambda_{1\to k}+\mu+2\theta) & \lambda_{1\to k}g^{1\to k}_1 & \lambda_{1\to k}g^{1\to k}_2 & \ldots
	\\
	0 & 0 & \mu+3\theta & -(\lambda_{1\to k}+\mu+3\theta) & \lambda_{1\to k}g^{1\to k}_1 & \ldots
	\\
	\vdots & \vdots & \ddots & \ddots & \ddots & \ddots
	\end{array}
	\right).
\end{equation}

The stationary distribution $\pi^{1\to k}=(\pi^{1\to k}_n)_{n\in\mathbb N}$ of $L^{1\to k}$ hence satisfies the following system of equations:
\begin{equation}
	\left\{\begin{array}{rl}
	0 &= -\lambda_{1\to k} \pi^{1\to k}_0 + (\mu+\theta)\pi^{1\to k}_1,
	\\
	0 &= -(\lambda_{1\to k}+\mu+n\theta) \pi^{1\to k}_n + (\mu+(n+1)\theta)\pi^{1\to k}_{n+1}
	+ \sum_{i=1}^n \lambda_{1\to k}g^{1\to k}_i \pi^{1\to k}_{n-i}, \;\; (n\geq 1),	
	\end{array}\right.
\end{equation}
which can be solved by introducing the probability generating functions. Let $\Pi^{1\to k}(z)=\sum_{n\in\mathbb N} \pi^{1\to k}_n z^n$ and $G^{1\to k}(z)=\sum_{n\in\mathbb N^*} g^{1\to k}_n z^n$. Let us also introduce the normalized parameters 
\begin{equation}
	\delta=\frac{\mu}{\theta} \;\; \text{ and } \;\; \nu_{1\to k}=\frac{\lambda_{1\to k}}{\theta}.
\end{equation}
By multiplicating the $n$-th line by $z^n$ and summing, the previous system leads to the following differential equation:
\begin{equation}
	\label{eq:PiDifferentialEquation}
	\frac{d}{dz}\Pi^{1\to k}(z)+\left(\frac{\delta}{z}-\nu_{1\to k}H^{1\to k}(z)\right) \Pi^{1\to k}(z) = \frac{\delta}{z} \pi^{1\to k}_0,
\end{equation}
where we have set $H^{1\to k}(z)=\frac{1-G^{1\to k}(z)}{1-z}$.
This equation is straightforwardly solved to obtain:
\begin{equation}
	\Pi^{1\to k}(z) = z^{-\delta} \delta \pi^{1\to k}_0 e^{\nu_{1\to k}\int_0^z H^{1\to k}(u)\,du} \int_0^z v^{\delta-1} e^{-\nu_{1\to k}\int_0^v H^{1\to k}(u)\,du} \,dv,
\end{equation}
and the condition $\Pi^{1\to k}(1)=1$ leads to
\begin{equation}
	\pi^{1\to k}_0 = \left(\delta \int_0^1 v^{\delta-1} e^{\nu_{1\to k}\int_v^1 H^{1\to k}(u)\,du} \,dv\right)^{-1},
\end{equation}
which by substituting in the general solution gives:
\begin{equation}
	\Pi^{1\to k}(z) = z^{-\delta} \frac{\int_0^z v^{\delta-1} e^{\nu_{1\to k}\int_v^z H^{1\to k}(u)\,du} \,dv}{\int_0^1 v^{\delta-1} e^{\nu_{1\to k}\int_v^1 H^{1\to k}(u)\,du} \,dv}.
\end{equation}
Now, turning back to the differential equation \eqref{eq:PiDifferentialEquation}, then taking the limit when $z$ tends increasingly to $1$ and using basic properties of probability generating function ($\lim_{\substack{z\to 1\\t<1}}\Pi^{1\to k}(z)=1$, $\lim_{\substack{z\to 1\\t<1}}\frac{d}{dz}\Pi^{1\to k}(z)=\mathbf E[L^{1\to k}]$ and $\lim_{\substack{z\to 1\\t<1}} H^{1\to k}(z) = \overline{g^{1\to k}}$), we obtain the result stated in the following proposition.

\begin{proposition}
In the discrete one-side order book model with Poisson arrival at rate $\mu$ of unit size market orders, Poisson arrival of limit orders with rate $\lambda_k$ at price $k$ and random size with distribution $(g^k_n)_{n\in\mathbb N^*}$ on $\mathbb N^*$, and exponential lifetime of non-executed limit orders with parameters $\theta$, 
the average cumulative shape of the order book up to price $k$ is given by:
\begin{equation}
	\label{eq:EL1tokVolume}
	\mathbf E[L^{1\to k}] = \nu_{1\to k} \overline{g^{1\to k}} - \delta + \left( \int_0^1 v^{\delta-1} e^{\nu_{1\to k}\int_v^1 H^{1\to k}(u)\,du} \,dv\right)^{-1}.
\end{equation}
\end{proposition}

Note that by taking the sizes of all limit orders to be equal to $1$, i.e. by setting $g^k_1=1$ and $g^k_n=0, n\geq 2$, equation \eqref{eq:EL1tokVolume} reduces to equation \eqref{eq:DiscreteCumulativeShape} of section \ref{sec:OBQueue}, as expected.

We now introduce a specification of the model where the sizes of limit orders are geometrically distributed with parameter $q\in(0,1)$ and independent of the price, i.e. for any price $k\in\mathbb N^*$, $g^k_n=(1-q)^{n-1}q$. 
This specification is empirically founded, as it has been observed that the exponential distribution may be a rough continuous approximation of the distribution of the sizes of limit orders \citep[see e.g.][]{Chakraborti2011}.
Since the distribution is independent of the price, then for any price $k\in\mathbb N^*$, $g^{1\to k}_n=(1-q)^{n-1}q$. This straightfowardly gives $H^{1\to k}(z)=\frac{1}{1-(1-q)z}$ and with some computations we obtain:
\begin{equation}
	\mathbf E[L^{1\to k}] = \frac{\nu_{1\to k}}{q} - \delta + \frac{\delta q^{\frac{\nu_{1\to k}}{1-q}}}
	{_2F_1(\delta,\frac{-\nu_{1\to k}}{1-q},1+\delta,1-q)},
\end{equation}
where $_2F_1$ is the ordinary hypergeometric function \citep[see e.g.][chapter 2]{Seaborn1991}.

Now, following the idea presented in section \ref{sec:OBQueue}, we consider an order book with a continuous price, in which limit orders ar submitted according to a spatial Poisson process with intensity $\lambda(p,t)=\alpha h_\lambda(p)$. Recall that $h_\lambda$  is assumed to be a real non-negative function with positive support, denoting the spatial intensity of arrival rates, i.e. the function such that the number of limit orders submitted in the order book in the price interval $[p_1,p_2]$ is a homogeneous Poisson process with rate $\int_{p_1}^{p_2}\alpha h_\lambda(u)\,du$.
Using notations defined in section  \ref{sec:OBQueue}, the cumulative shape at price $p\in[0,+\infty)$ of this continuous order book is thus:
\begin{equation}
	\label{eq:ContinuousOBCumShapeGeometricSizes}
	B(p) = \frac{1}{q}H(p)- \delta + \frac{\delta q^{\frac{H(p)}{1-q}}}
	{_2F_1(\delta,\frac{-H(p)}{1-q},1+\delta,1-q)},
\end{equation}
which can be derived to obtain the average shape $b(p)$ of the order book, which we state in the following proposition.
\begin{proposition}
In a continuous one-side order book with homogeneous Poisson arrival of unit-size market orders with intensity $\mu$, spatial Poisson arrival of limit orders intensity $\alpha h_\lambda(p)$, geometric distribution of the sizes of limit orders with parameter $q$, and exponentially distributed lifetimes of non-executed limit orders with parameter $\theta$, the average shape of the order book $b$ is computed for all $p\in[0,\infty)$ by:
\begin{equation}
	\label{eq:ContinuousOBShapeGeometricSizes}
	b(p) = \frac{h(p)}{q} + \frac{d}{dp}\left(\frac{\delta q^{\frac{H(p)}{1-q}}}
	{_2F_1(\delta,\frac{-H(p)}{1-q},1+\delta,1-q)}\right).
\end{equation}
\end{proposition}

\section{Influence on the size of limit orders on the shape of the order book}
\label{sec:InfluenceOrderSize}
We now use the results of section \ref{sec:ModelVolumes} to investigate the influence of the size of the limit orders on the shape of the order book. Recall that market orders are submitted at rate $\mu$ with size $1$, that non-executed limit orders are cancelled share by share after a random time with exponential distribution with parameter $\theta$, and that the distribution of the sizes of limit orders is a geometric distribution with parameter $q$ (i.e. with mean $\frac{1}{q}$). The first subsection details the influence of the parameter $q$ on the theoretical shape of the order book. The second one provides empirical evidence confirming the findings of our model.

\subsection{The order book shape as a function of the average size of limit orders}
\begin{figure}
\begin{tabular}{c}
\includegraphics[scale=1]{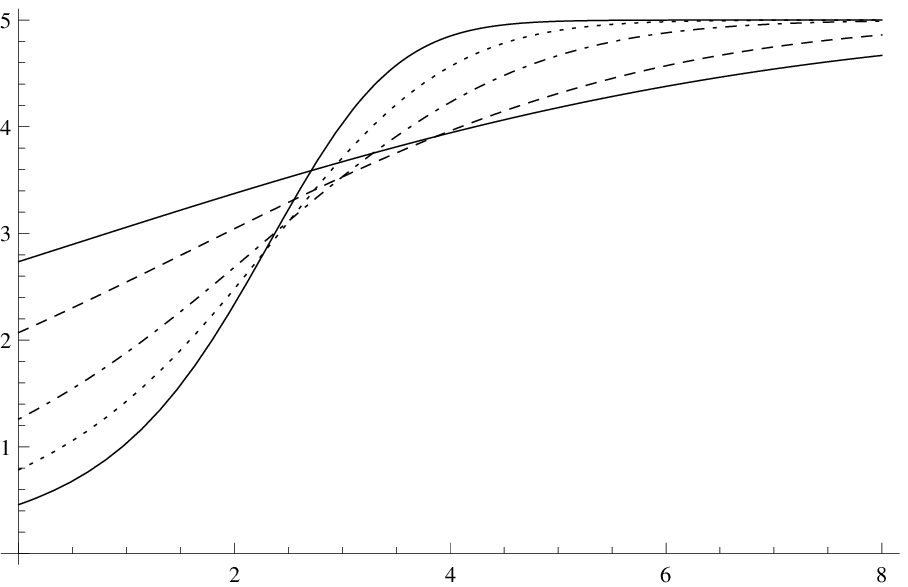}
\\
\includegraphics[scale=1]{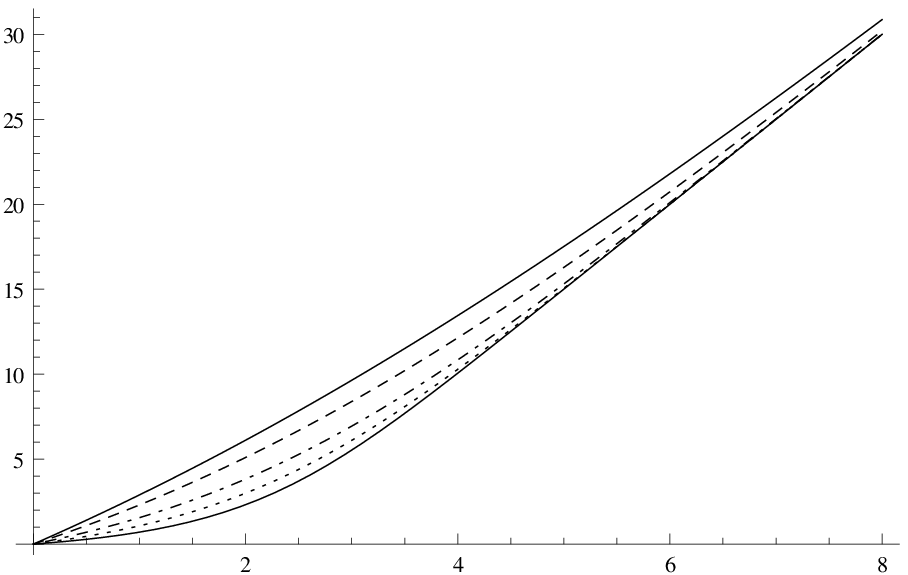}
\end{tabular}
\caption{Shape of the order book as computed in equation \eqref{eq:ContinuousOBShapeGeometricSizes} (top) and cumulative shape of the order book as computed in equation \eqref{eq:ContinuousOBCumShapeGeometricSizes} (bottom) with $\delta=10$, $h(p)=\frac{\alpha}{qK}\mathbf 1_{(0,K)}$, $\alpha=40$, $K=8$, and $q=0.99$ (lower full line at $0$), $q=0.5$ (dotted), $q=0.25$ (dotdashed), $q=0.10$ (dashed), $q=0.05$ (upper full line at $0$)}
\label{fig:OBShapeConstantLambdaVarq}
\end{figure}

In a first example, we assume that the normalized intensity of arrival of limit orders $h$ is constant \citep[i.e. as in][]{Smith2003} and equal to $\frac{\alpha}{q}$.
Note that when $q$ varies, the mean total volume $\overline{V}(p)$ of arriving limit orders up to price $p$ per unit time remains constant:
\begin{equation}
	\label{eq:CstArrivingVolumeLimitOrders}
	\overline{V}(p) = q\int_0^p h(u)\,du = p\alpha.
\end{equation}
In other words, when $q$ decreases, limit orders are submitted with larger sizes in average, but less often, keeping the total submitted volumes constant.
The first remarkable observation is that, although the mean total volumes of limit and market orders are constant, the shape of the order book varies widely with $q$.
On figure \ref{fig:OBShapeConstantLambdaVarq}, we plot the shape $b(p)$ defined at equation \eqref{eq:ContinuousOBShapeGeometricSizes}, and cumulative shape defined at equation \eqref{eq:ContinuousOBCumShapeGeometricSizes}, of an order book with arriving volumes of limit orders as in equation \eqref{eq:CstArrivingVolumeLimitOrders}. With the chosen numerical values, the average volume of one limit order ranges from approximately $1$ $(q=0.99)$ to $20$ $(q=0.05)$. 
It appears that when $q$ decreases, the shape of the order book increases for lower prices. In other words, \emph{the larger the size of arriving limit orders is, the deeper is the order book around the spread}, all other things being equal.

We observe that figure \ref{fig:OBShapeConstantLambdaVarq} here is similar to figure \ref{fig:OBShapeCumShapeUnitVolumeVaryingDelta} here and figure 3 in \cite{Smith2003}.
However, volumes of limit and market orders are equal in the two latter cases, and we have shown in section \ref{sec:Comparison} that these different shapes can actually be obtained with different regimes of market orders, but equal sizes of market and limit orders: when the arrival rates of market orders increases, all other things being equal, the average shape of the order book is thinner for lower prices. This observation does not depend on the relative volumes of limit orders and market orders.
Therefore, the observation made now is different. In \emph{similar trading regimes} where the mean total volume of limit and market orders are equal, we highlight the influence of the relative volume of limit orders (compared to unit market orders) on the order book shape: the smaller the average size of limit orders, the shallower the order book close to the spread.

\begin{figure}
\begin{center}
\includegraphics[scale=1]{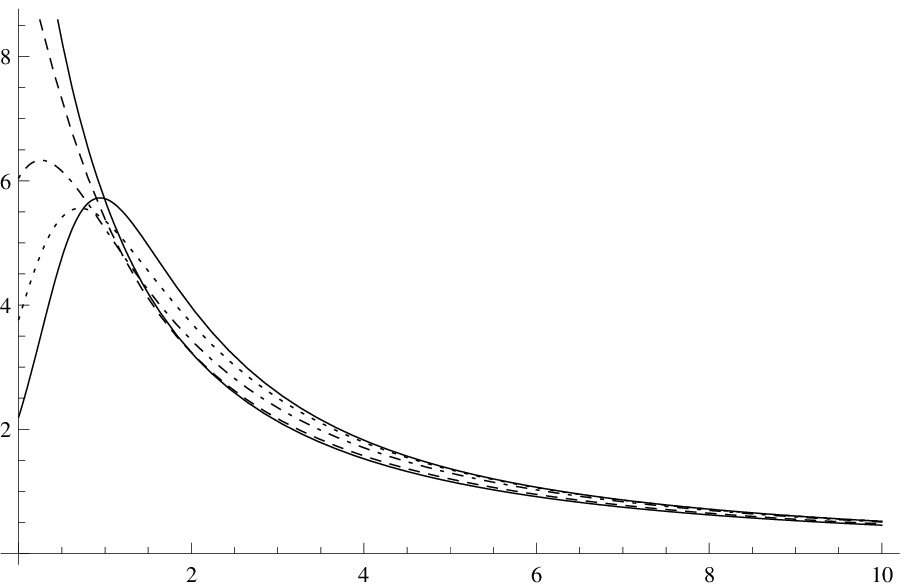}
\end{center}
\caption{Shape of the order book as computed in equation \eqref{eq:ContinuousOBShapeGeometricSizes} with $\delta=10$, $h(p)=q^{-1}\alpha (\gamma-1)(1+p)^{-\gamma}$, $\alpha=40$ $\gamma=1.6$, and $q=0.99$ (lower full line at $0$), $q=0.5$ (dotted), $q=0.25$ (dotdashed), $q=0.10$ (dashed), $q=0.05$ (upper full line at $0$).}
\label{fig:OBShapePowerLambdaVarq}
\end{figure}

We provide a second exemple of the phenomenon by assuming that the intensity of incoming limit orders exhibits a power-law decrease with the price, as tested in section \ref{sec:Comparison} to compare to the analytical shape provided in \cite{Bouchaud2002}. We thus have now $h(p) = q^{-1}\alpha (\gamma-1)(1+p)^{-\gamma}$. There again, when $q$ varies, the average total volume of incoming limit orders up to price $p$ remains constant and equal to $\alpha (\gamma-1) \int_0^p(1+u)^{-\gamma}\,du = \alpha (1-(1+p)^{1-\gamma})$. Figure \ref{fig:OBShapePowerLambdaVarq} plots the shape of the order book with these characteristics, when $q$ varies.
The observed phenomenon is equally clear with this more realistic distribution of incoming limit orders: the order book deepens at the first limits when the average volume of limit orders increases, the total volume of limit orders and unit-size market orders submitted being constant.

Finally, one might remark that, by assuming unit-size market orders, our model with geometric distribution of limit orders' sizes does not predict the order book shape in the case where the average size of limit orders is smaller than the average size of market orders. In fact, it will now appear that this case has never been empirically encountered in our data: the average size of limit orders is always in our sample greater than the average size of market orders.

\subsection{Empirical results}
We end this paper with an empirical study providing some evidence of the previous results. We use Thom\-son-Reu\-ters tick-by-tick data for fourteen stocks traded on the Paris stock exchange, from January 4th, 2010 to February 22nd, 2010. The fourteen stocks under investigation are: Air Liquide (AIRP.PA, chemicals), Alstom (ALSO.PA, transport and energy), Axa (AXAF.PA, insurance), BNP Paribas (BNPP.PA, banking), Bouygues (BOUY.PA, construction, telecom and media), Carrefour (CARR.PA, retail distribution), Danone (DANO.PA, milk and cereal products), Lagardère (LAGA.PA, media), Michelin (MICP.PA, tires manufacturing), Peugeot (PEUP.PA, vehicles manufacturing), Renault (RENA.PA, vehicles manufacturing), Sanofi (SASY.PA, healthcare), Vinci (SGEF.PA, construction and engineering), Ubisoft (UBIP.PA, video games).
All these stocks except Ubisoft were included in the CAC 40 French index in January and February 2010, i.e. they are among the largest market capitalizations and most liquid stocks on the Paris stock exchange (see descriptive statistics below).

All movements on the first $10$ limits of the ask side and the bid side of the order book are available, which allows us to reconstruct the evolution of the first limits of the order book during the day.
Each trading day is divided into $12$ thirty-minute intervals from $10$am to $4$pm. We obtain $T=391$ intervals for each stock\footnote{Three and a half days of trading are missing or with incomplete information in our dataset: January 15th, the morning of January 21st, February 18th and 19th.}.
For each interval $t=1,\ldots,T$, and for each stock $k=1,\ldots,14$, we compute the total number of limit orders $N^{\lambda}_{k,t}$ and market orders $N^{\mu}_{k,t}$ and the average sizes of limit orders $V^{\lambda}_{k,t}$ and market orders $V^{\mu}_{k,t}$.
Table \ref{table:descriptiveStatistics} gives the average number of orders and their volumes (the overline denotes the average over the time intervals: $\overline{N}^{\mu}_k = \frac{1}{T}\sum_{t=1}^T N^{\mu}_{k,t}$, and similarly for other quantities).
The lowest average activity is observed on UBIP.PA and LAGA.PA (which are the only two stocks in the sample with less than $200$ market orders and $4000$ limit orders in average). The highest activity is observed on BNPP.PA (which is the only stock with more than $600$ market orders and $6000$ limit orders in average). The smallest sizes of orders are observed on AIRP.PA ($108.25$ and $205.4$ for market and limit orders), and the largest sizes are observed on AXA.PA ($535.6$ and $876.6$ for market and limit orders).
\begin{table}
\footnotesize
\begin{center}
\begin{tabular}{|c|cc|cc|cc|cc|}%cc|}
\hline
Stock $k$
& \multicolumn{2}{|c|}{$\overline{N^{\mu}_k}$} & \multicolumn{2}{|c|}{$\overline{V^{\mu}_k}$} & \multicolumn{2}{|c|}{$\overline{N^{\lambda}_k}$} & \multicolumn{2}{|c|}{$\overline{V^{\lambda}_k}$} %& \multicolumn{2}{|c|}{$\overline{S}_k$}
\\ 
& ($\min$, & $\max$) & ($\min$, & $\max$) & ($\min$, & $\max$) & ($\min$, & $\max$) % & ($\min$, & $\max$) 
\\ \hline
AIRP.PA & \multicolumn{2}{|c|}{290.9}  &  \multicolumn{2}{|c|}{108.2}  &  \multicolumn{2}{|c|}{4545.4}  &  \multicolumn{2}{|c|}{205.4} %&  \multicolumn{2}{|c|}{4.82}
\\ &  59  &  1015  &  60.2  &  236.9  &  972  &  24999  &  153.6  &  295.7 %&  2.30  &  8.62
\\
ALSO.PA & \multicolumn{2}{|c|}{349.7}  &  \multicolumn{2}{|c|}{181.3}  &  \multicolumn{2}{|c|}{5304.2}  &  \multicolumn{2}{|c|}{289.2} %& \multicolumn{2}{|c|}{6.08}
\\ &  68  &  1343  &  75.7  &  310.8  &  817  &  31861  &  234.5  &  389.6 %&  3.18  &  9.53
\\
AXAF.PA & \multicolumn{2}{|c|}{521.8}  &  \multicolumn{2}{|c|}{535.6}  &  \multicolumn{2}{|c|}{4560.6}  &  \multicolumn{2}{|c|}{876.6} %& \multicolumn{2}{|c|}{1.97} 
\\ &  119  &  2169  &  311.4  &  1037.4  &  1162  &  20963  &  603.4  &  1648.3 %&  1.15  &  3.85 
\\ 
BNPP.PA & \multicolumn{2}{|c|}{773.7}  &  \multicolumn{2}{|c|}{203.1}  &  \multicolumn{2}{|c|}{6586.0}  &  \multicolumn{2}{|c|}{277.6} %& \multicolumn{2}{|c|}{2.73} 
\\ &  160  &  4326  &  113.1  &  379.5  &  946  &  42939  &  189.2  &  518.7 %&  1.09  &  4.97
\\
BOUY.PA & \multicolumn{2}{|c|}{218.3}  &  \multicolumn{2}{|c|}{227.4}  &  \multicolumn{2}{|c|}{4544.5}  &  \multicolumn{2}{|c|}{369.7} %& \multicolumn{2}{|c|}{5.33} 
\\ &  38  &  1021  &  113.1  &  421.9  &  652  &  25546  &  274.5  &  475.2 %&  2.22  &  8.65
\\
CARR.PA & \multicolumn{2}{|c|}{309.5}  &  \multicolumn{2}{|c|}{275.3}  &  \multicolumn{2}{|c|}{4391.4}  &  \multicolumn{2}{|c|}{513.3} %& \multicolumn{2}{|c|}{3.90} 
\\ &  49  &  1200  &  156.1  &  517.3  &  813  &  14752  &  380.8  &  879.6 %&  1.60  &  7.96
\\
DANO.PA & \multicolumn{2}{|c|}{385.7}  &  \multicolumn{2}{|c|}{214.4}  &  \multicolumn{2}{|c|}{4922.1}  &  \multicolumn{2}{|c|}{393.1} %& \multicolumn{2}{|c|}{3.54} 
\\ &  86  &  2393  &  112.6  &  820.3  &  1372  &  18186  &  286.1  &  537.4 %&  1.83  &  6.87
\\
LAGA.PA & \multicolumn{2}{|c|}{140.8}  &  \multicolumn{2}{|c|}{201.5}  &  \multicolumn{2}{|c|}{3429.0}  &  \multicolumn{2}{|c|}{338.2} %& \multicolumn{2}{|c|}{4.37} 
\\ &  22  &  544  &  87.2  &  376.8  &  632  &  11319  &  219.0  &  504.0 %&  2.01  &  7.37
\\ 
MICP.PA & \multicolumn{2}{|c|}{301.2}  &  \multicolumn{2}{|c|}{137.9}  &  \multicolumn{2}{|c|}{5033.5}  &  \multicolumn{2}{|c|}{240.7} %& \multicolumn{2}{|c|}{4.23} 
\\ &  45  &  1114  &  74.4  &  235.4  &  1012  &  23799  &  178.0  &  315.1 %&  2.34  &  6.63
\\
PEUP.PA & \multicolumn{2}{|c|}{294.9}  &  \multicolumn{2}{|c|}{333.2}  &  \multicolumn{2}{|c|}{3790.6}  &  \multicolumn{2}{|c|}{536.7} %& \multicolumn{2}{|c|}{4.42} 
\\ &  57  &  1170  &  159.1  &  657.7  &  967  &  14053  &  316.8  &  828.3 %&  1.95  &  7.29
\\
RENA.PA & \multicolumn{2}{|c|}{463.2}  &  \multicolumn{2}{|c|}{266.9}  &  \multicolumn{2}{|c|}{4957.6}  &  \multicolumn{2}{|c|}{384.3} %& \multicolumn{2}{|c|}{5.34} 
\\ &  103  &  1500  &  133.7  &  477.2  &  1383  &  27851  &  283.1  &  539.8 %&  2.16  &  8.79
\\
SASY.PA & \multicolumn{2}{|c|}{494.0}  &  \multicolumn{2}{|c|}{241.4}  &  \multicolumn{2}{|c|}{4749.1}  &  \multicolumn{2}{|c|}{421.2} %& \multicolumn{2}{|c|}{2.40} 
\\ &  106  &  1722  &  128.7  &  511.7  &  986  &  22373  &  311.1  &  670.2  %&  1.31  &  4.39
\\ 
SGEF.PA & \multicolumn{2}{|c|}{366.1}  &  \multicolumn{2}{|c|}{188.4}  &  \multicolumn{2}{|c|}{5309.9}  &  \multicolumn{2}{|c|}{353.4} %& \multicolumn{2}{|c|}{4.57} 
\\ &  70  &  1373  &  107.7  &  452.3  &  983  &  21372  &  274.2  &  515.1 %&  2.00  &  7.53
\\
UBIP.PA & \multicolumn{2}{|c|}{153.3}  &  \multicolumn{2}{|c|}{384.9}  &  \multicolumn{2}{|c|}{1288.9}  &  \multicolumn{2}{|c|}{754.0} %& \multicolumn{2}{|c|}{2.44} 
\\ &  18  &  998  &  198.4  &  675.1  &  240  &  7078  &  451.4  &  1121.0 %& 0.93 &  4.95
\\ \hline
\end{tabular}
\end{center}
\caption{Basic statistics on the number of orders and the average volumes of orders per $30$-minute time interval for each stock.} 
%Spreads are given in ticks.}
\label{table:descriptiveStatistics}
\end{table}

\begin{figure}
\begin{center}
\includegraphics[scale=0.5]{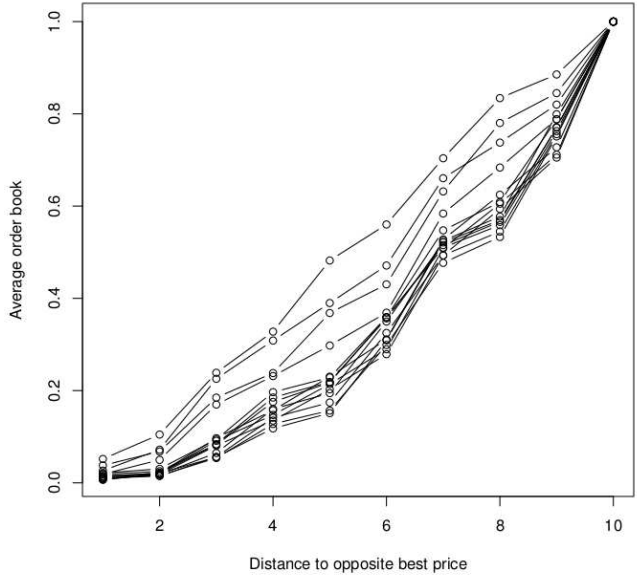}
\caption{Mean scaled shapes of the cumulative order book up to $10$ ticks away from the opposite best price for the 14 stocks studied.}
\end{center}
\label{fig:EmpiricalMeanOBShapes}
\end{figure}

We also compute the average cumulative order book shape at $1$ to $10$ ticks from the best opposite side. The average cumulative depth $B^i_{k,t}$ is thus the quantity available in the order book for stock $k$ in the price range $\{b(t)+1,\ldots,b(t)+i\}$ (in ticks) for ask limit orders, or in the price range $\{a(t)-i,\ldots,a(t)-1\}$ for bid limit orders, averaged over time during interval $t$.
For $i$ lower or equal to $10$, our data is always complete since the first ten limits are available. For larger $i$ however, we may not have the full data: $B^{10+j}_{k,t}$ is not exact if the spread reaches a level lower or equal to $j$ ticks during interval $t$. Hence $i$ is lower or equal to $10$ in the following empirical analysis.
Figure \ref{fig:EmpiricalMeanOBShapes} plots the empirical average shapes $\overline{B}^{i}_{k} = \frac{1}{T}\sum_{t=1}^T B^i_{k,t}$ (arbitrarily scaled to $\overline{B}^{10}_{k}=1$ for easier comparison).

Following our theoretical results from the previous sections, we investigate the influence of the number of limit orders $N^\lambda_{k,t}$ and their average size $V^\lambda_{k,t}$ on the depth on the order book at the first limits. The theoretical insight of the previous section gives a negative relationship with the number of limit orders and a positive relationship with the size of these orders, all others things being equal, i.e. the global market activity being held constant.
A fairly natural proxy for the market activity is the traded volume (transactions) per period. The total volume of market orders submitted during interval $t$ for the stock $k$ is $N^{\mu}_{k,t}V^{\mu}_{k,t}$. We may also consider the total volume of incoming orders  $N^\lambda_{k,t}V^\lambda_{k,t}$.

Therefore, we have the following regression model for some book depth $B_{k,t}$:
\begin{equation}
	\label{eq:EmpiricalModelEquation}
	B_{k,t} = \beta_1 N^\lambda_{k,t} + \beta_2 V^\lambda_{k,t} +
		\beta_3 N^\lambda_{k,t}V^\lambda_{k,t} + \beta_4 N^\mu_{k,t}V^\mu_{k,t} + \epsilon_{k,t}.
\end{equation}
We test this regression using $B_{k,t}=B^{5}_{k,t}$ (cumulative depth up to 5 ticks away from the best opposite price), then using $B_{k,t}=B^{10}_{k,t}$ (cumulative depth up to 10 ticks away from the best opposite price) and finally using $B_{k,t}=B^{10}_{k,t}-B^5_{k,t}$ (cumulative volume between 6 and 10 ticks away from the best opposite price). For each of these three models, we provide results with or without the interaction term $N^\lambda_{k,t}V^\lambda_{k,t}$.
All models are estimated as panel regression models with fixed effects, i.e. the error term $\epsilon_{k,t}$ is the sum of a non-random stock specific component $\delta_k$ (fixed effect) and a random component $\eta_{k,t}$. While the regression coefficients are stock-independent, the variables $\delta_k$ translate the idiosyncratic characteristics of each stock.
Results are given in table \ref{table:PanelRegressionResults}.
\begin{sidewaystable}
\footnotesize
\begin{center}
\begin{tabular}{lrrrrcrrrrc}
$B^{5}_{k,t}$
 & \multicolumn{5}{l}{Without $N^\lambda_{k,t}V^\lambda_{k,t}$}  
 & \multicolumn{5}{l}{With $N^\lambda_{k,t}V^\lambda_{k,t}$}
\\ Variables & Est. & Std Err. & $t$-value & $p$-value & Sig.
& Est. & Std Err. & $t$-value & $p$-value & Sig.
\\ \hline
$N^\lambda_{k,t}$ 
& -0.24455434  & 0.02008400 & -12.177 & $<$ 2.2e-16 & ***
& -8.6511e-02  & 3.6046e-02 & -2.4000 &  0.01643 & *  
\\
$V^\lambda_{k,t}$ 
& 8.70817318  & 0.75084050  & 11.598 & $<$ 2.2e-16 & ***
& 1.0374e+01  & 8.1288e-01 & 12.7622 & $<$ 2.2e-16 & ***
\\
$N^\lambda_{k,t}V^\lambda_{k,t}$ 
& --- & --- & --- & --- & ---
& 1.0794e-02 & 8.7448e-04 & 12.3430 & $<$ 2.2e-16 & ***
\\
$N^\mu_{k,t}V^\mu_{k,t}$ 
& 0.00882725  & 0.00079296  & 11.132 & $<$ 2.2e-16 & ***
& -4.8483e-04  & 9.1924e-05 & -5.2743 & 1.384e-07 & ***
\\ \hline
$R^2$ & 0.079275 & & & & & 0.083945
\\
$F$-statistic & 156.617 & & & & *** & 124.994 & & & & ***
\\ \hline
\hline
$B^{10}_{k,t}$
 & \multicolumn{5}{l}{Without $N^\lambda_{k,t}V^\lambda_{k,t}$}  
 & \multicolumn{5}{l}{With $N^\lambda_{k,t}V^\lambda_{k,t}$}
\\ Variables & Est. & Std Err. & $t$-value & $p$-value & Sig.
& Est. & Std Err. & $t$-value & $p$-value & Sig.
\\ \hline
$N^\lambda_{k,t}$ 
& -0.4576647 & 0.0439167 & -10.4212 & $<$ 2.2e-16 & ***
& -0.25599592 & 0.07895182 & -3.2424  & 0.001192 & ** 
\\
$V^\lambda_{k,t}$ 
& 39.4968315 &  1.6418268 & 24.0566 & $<$ 2.2e-16 & ***
& 41.62260408 & 1.78046702 & 23.3774& $<$ 2.2e-16 & ***
\\
$N^\lambda_{k,t}V^\lambda_{k,t}$ 
& --- & --- & --- & --- & ---
& 0.01222198 & 0.00191540 & 6.3809 & 1.906e-10 & ***
\\
$N^\mu_{k,t}V^\mu_{k,t}$ 
& 0.0097127 & 0.0017339 &  5.6016 &  2.228e-08 & ***
& -0.00061866 & 0.00020134 & -3.0727 & 0.002132 & ** 
\\ \hline
$R^2$ & 0.14202 & & & & & 0.1435 
\\
$F$-statistic & 301.1 & & & $<$ 2.2e-16 & *** & 228.534 & & & $<$ 2.2e-16 & ***
\\ \hline
\hline
$B^{10}_{k,t}-B^{5}_{k,t}$
 & \multicolumn{5}{l}{Without $N^\lambda_{k,t}V^\lambda_{k,t}$}  
 & \multicolumn{5}{l}{With $N^\lambda_{k,t}V^\lambda_{k,t}$}
\\ Variables & Est. & Std Err. & $t$-value & $p$-value & Sig.
& Est. & Std Err. & $t$-value & $p$-value & Sig.
\\ \hline
$N^\lambda_{k,t}$ 
& -0.21311036 & 0.02995514 & -7.1143 & 1.27e-12 & ***
& -0.16948494 & 0.05389411 & -3.1448 & 0.001671 & ** 
\\
$V^\lambda_{k,t}$ 
& 30.78865834 & 1.11987302 & 27.4930 & $<$ 2.2e-16 & ***
& 31.24851002 & 1.21538277 & 25.7108 & $<$ 2.2e-16 & ***
\\
$N^\lambda_{k,t}V^\lambda_{k,t}$ 
& --- & --- & --- & --- & ---
& 0.00142825 & 0.00130749 & 1.0924 & 0.274723    
\\
$N^\mu_{k,t}V^\mu_{k,t}$ 
& 0.00088543 & 0.00118269 & 0.7487  &  0.4541 &   
& -0.00013383 & 0.00013744 & -0.9737 & 0.330234 &   
\\ \hline
$R^2$ & 0.14925 & & & & & 0.1435 
\\
$F$-statistic & 318.743 & & & $<$ 2.2e-16 & *** & 239.292 & & & $<$ 2.2e-16 & ***
\\ \hline
\end{tabular}
\end{center}
\caption{Panel regression results for the models defined in equation \eqref{eq:EmpiricalModelEquation}, for $B_{k,t}=B^{5}_{k,t}$ (top panel), $B_{k,t}=B^{10}_{k,t}$ (middle panel) and $B_{k,t}=B^{10}_{k,t}-B^{5}_{k,t}$ (lower panel). In all case, data has $14$ stocks and $T=391$ time intervals, i.e. $5474$ points.}
\label{table:PanelRegressionResults}
\end{sidewaystable}

These empirical results are in accordance with both previous empirical observations and the new insights from the previous sections.
First of all, we observe a globally positive relationship linking the cumulative order book depth (up to $5$ or $10$ ticks away from the opposite best price) and the global market activity. 
If the global market activity is only represented by the transactions $N^{\mu}_{k,t}V^{\mu}_{k,t}$, then there is indeed a positive relationship between the depth of the order book and $N^{\mu}_{k,t}V^{\mu}_{k,t}$. This is in accordance with existing financial observations, since it is known that high market activity tends to "tighten" the spread and increase the volumes available around the spread. Such a phenomenon is for example observed in \cite{Naes2006}, where a positive relationship between the number of trades and the order book slope of the first $5$ limits of the order book is exhibited. This situation is however somewhat paradoxical, since market orders should mechanically reduce the depth of the order book at the first limits.
Our empirical results clear the apparent paradox of the standard observations: if the volumes of both market and limit orders are included in the model, then we observe that this positive effect of market activity is in fact due to the volume of submitted limit orders $N^{\lambda}_{k,t}V^{\lambda}_{k,t}$, while 
the effect of the market orders $N^{\mu}_{k,t}V^{\mu}_{k,t}$ is actually negative (but with a smaller amplitude), as expected. 
Another important comment deals with the significance of the relationship between global market activity and order book depth. It is also observed on table \ref{table:PanelRegressionResults} that this relationship is not significant anymore if we take only the "furthest" limits, i.e. the limits between 6 and 10 ticks away from the best price.
It is very interesting to remark that \cite{Naes2006} also find that this positive relation between order book depth and market activity is much stronger closer to the spread, and then decresases when taking into account further limits.

A second series of results deals with the coefficients of the terms $N^\lambda_{k,t}$ and $V^\lambda_{k,t}$ separately. There is indeed a positive relationship between the order book depth and the average size of limit orders $V^\lambda_{k,t}$, but a negative relationship between the order book depth and the number of limit orders $N^\lambda_{k,t}$. Note also that the estimated $\beta$'s for these two quantities are all significant to the $0.1\%$ level without the interaction term $N^\lambda_{k,t}V^\lambda_{k,t}$ or $2\%$ (at worst) with this term. 
The first fact seems very intuitive: the larger the arriving orders, the deeper the order book. The second one is not as intuitive: at first glance, the more arriving limit orders, the deeper should be the order book. All others things being equal, it thus appears we have identified the effect we've just described using the theoretical model of section \ref{sec:ModelVolumes}: for a given total volume of arriving limit and market orders $N^\lambda_{k,t}V^\lambda_{k,t}$ and $N^\mu_{k,t}V^\mu_{k,t}$, the relative size of the limit orders has a strong influence on the average shape of the order book. The average shape of the order book is deeper when a few large limit orders are submitted, than when many small limit orders are submitted.

\section{Conclusion}
\label{sec:Conclusion}
We have introduced in this paper a simple order book model based on classic results from queueing theory. We have provided a continuous version of the model and shown that it provides an analytical formula for the shape of the order book that reproduces results from acknowledged numerical and empirical studies. We then have extended the model to allow for the limit orders to be submitted with random sizes. The extended model provides hints on the influence of the size of limit orders in an order book. We have confirmed these hints empirically by a study on liquid stocks traded on the Paris Stock Exchange.

We hope this theoretical and empirical study will encourage the use of queueing theory results in the field of order book modelling. We are confident that more results should follow using similar techniques.

\bibliographystyle{authordate1}
\bibliography{TheOrderBookAsAQueueingSystem}

\end{document}